\documentclass[12pt]{article}
\usepackage{graphicx}

\newsavebox{\sboxpubnumber}
\newsavebox{\sboxpubdate}
\newcommand{\pubdate}[1]{\begin{lrbox}{\sboxpubdate}{#1}\end{lrbox}}

\newcommand{\Title}[1]{\begin{center} {\Large #1 } \end{center}}
\newcommand{\Author}[1]{\begin{center}{ \sc #1} \end{center}}
\newcommand{\Address}[1]{\begin{center}{ \it #1} \end{center}}

\newenvironment{Abstract}{\begin{quotation}  }{\end{quotation}}
\newenvironment{Presented}{\begin{quotation} \begin{center}
             PRESENTED AT\end{center}\bigskip
      \begin{center}\begin{large}}{\end{large}\end{center}
      \end{quotation}}
\newcommand{\Acknowledgements}{\bigskip  \bigskip \begin{center} \begin{large}
             \bf ACKNOWLEDGEMENTS \end{large}\end{center}}

\newcommand{\beq}{\begin{equation}}
\newcommand{\dd}{\partial}
\newcommand{\eeq}{\end{equation}}
\newcommand{\bea}{\begin{eqnarray}}
\newcommand{\eea}{\end{eqnarray}}

\begin{document}

\begin{titlepage}
\pubdate{November 29, 2001}                    

\vfill
\Title{Neutrino production in a time-dependent matter background}
\vfill
\Author{Marieke Postma}
\Address{Department of Physics and Astronomy, UCLA, Los Angeles, CA
90095-1547}

\vfill
\begin{Abstract}
We show that neutrinos can be produced through standard electroweak
interactions in matter with time-dependent density.
\end{Abstract}
\vfill
\begin{Presented}
    COSMO-01 \\
    Rovaniemi, Finland, \\
    August 29 -- September 4, 2001
\end{Presented}
\vfill
\end{titlepage}
\def\thefootnote{\fnsymbol{footnote}}
\setcounter{footnote}{0}


\section{Introduction}

In this talk I will show that neutrinos can be pair produced by a
time-dependent matter density through standard electroweak
interactions.  The idea is that both ordinary matter and the nuclear
matter found in neutron stars carry a net SU(2) charge.  Neutrinos
couple to this charge through the electroweak interactions. In a
time-dependent matter background this leads to neutrino pair
production~\cite{nu_prod}. This phenomenon is analagous to electron
pair production by a time-varying electromagnetic
field~\cite{ct,grib}, or fermion production by a variable scalar field
as can occur during preheating~\cite{baa,gk}. Applications include
density waves in a neutron star, a neutron star in a binary system, or
a supernova explosion. We expect the amount of neutrinos produced to
be small though.

\section{Formalism}

The starting point in our description of neutrino production is the
Dirac equation for a neutrino in the presence of a time-dependent
electroweak current $j^\mu(t)$:
\begin{equation}
\left [ i \gamma^0  \dd_0 + i \gamma^k \dd_k +j_\mu(t)  \gamma^\mu
P_{\rm L} -m
\right ]  \psi =0.
\end{equation} 
Here $P_{\rm L}$ is the projection operator on left handed states.  In
electrically neutral matter, the isospin of protons and electrons
cancel, and the net weak current is proportional to the number density
of neutrons $n_n$. Let us consider the case of a spatially uniform
matter density moving with a four-velocity $v_\mu$. Then $j_\mu=\rho
v_\mu$ with $\rho = (G_{F}/\sqrt{2})n_n$. The largest values of the
``reduced density'' $\rho$ are obtained in nuclear matter. In a
neutron star the reduced density ranges from $\rho \sim 10^{3} \,{\rm
eV}$ in core to $\rho \sim 10^{-3} \,{\rm eV}$ in the outer layers.

To solve the Dirac equation it proves convenient to rewrite it as a
second order differential equation for scalar functions.  To do so we
expand the neutrino field in a complete set of eigenmodes
\beq
	\psi = \sum_{s=\pm} \int \frac{d^3 k}{(2\pi)^3} \left [
	b_{{\bf k},s} U_{{\bf k},s} (t) + 
	d_{-{\bf k},s}^{\dag} V_{-{\bf k},s}(t) 
	\right ] e^{i {\bf k}{\bf x} }.
\label{exp}
\eeq 
For the eigenspinors we make the following {\it Ansatz}
\bea 
	U_{{\bf k},s} (t) e^{i {\bf k}{\bf x}} &=& 
	N_{{k},s} \left [ i \gamma^\mu
	\dd_\mu +m + j_\mu(t) \gamma^\mu P_{\rm L} \right ]
	\phi^{(1)}_{k,s}(t) R_{k,s}^{(1)}
	{\rm e}^{i {\bf k}{\bf x}-\i /2 \int_0^t \rho_\pm (\zeta) d
	\zeta }, \nonumber \\
	V_{{\bf -k},s} (t) e^{i {\bf k}{\bf x}} &=& 
	N_{{k},s} \left [ i \gamma^\mu
	\dd_\mu +m + j_\mu(t) \gamma^\mu P_{\rm L} \right ]
	\phi^{(-1)}_{k,s}(t) R_{k,s}^{(-1)}
	{\rm e}^{i {\bf k}{\bf x} -\i /2 \int_0^t \rho_\pm (\zeta) d
	 \zeta }.
\label{Ansatz}
\eea
$N_{{k},s}$ are normalization constants.  We choose the
time-independent spinors $R_{k,s}^{(\lambda)}$, $\lambda = \pm 1$, such
that they transform into each other under the action of $\gamma^0$,
and that they are eigenvectors of the spin operator:
\beq
	\gamma^0 R_{k,s}^{(\lambda)} = R_{k,s}^{(- \lambda)}, \qquad
	\gamma^0 \vec{\gamma}  \cdot \vec{k} R_{k,\pm}^{(\lambda)} 
	= \pm R_{k,\pm}^{(\lambda)}.
\eeq
This latter property means that for matter moving with some velocity,
we choose $\vec{v} \parallel \vec{k}$.  We do not expect the
orthogonal momenta to change the overall picture of particle
production.  Substituting this {\it Ansatz} back in the Dirac
equations gives the desired equation for the mode functions
$\phi^{(\lambda)}_{k,s}$:
\beq
	\ddot{\phi}_{k,\pm}^{(\lambda)} + \left [ m^2 +  
	k_{\rm eff,\pm}^2 - i \frac{\lambda}{2}
	\dot{\rho}_\pm \right ] \phi_{k,\pm}^{(\lambda)} = 0,
\label{mode}
\eeq 
where we have aligned the neutrino momentum and the velocity of the
background matter with the $3$rd axis, and we introduced the shorthand
notation $\rho_\pm = \rho_0(v_0 \pm v_3)$. Furthermore $k_{\rm
eff,\pm}(t)= \pm k + \rho_\pm(t)/2$ is the time-dependent effective
neutrino momentum. The shift in momentum is due to the presence of the
electroweak potential generated by the matter background.

The solutions corresponding to the two values of $\lambda$ satisfy
equations that are complex conjugate of each other, {\it i.e.},
\begin{equation}
\label{rel_phi}
\left (\phi_{k,\pm}^{(\lambda)} \right )^*= \phi_{k,\pm}^{(-\lambda)}. 
\label{cp}
\end{equation}
Hence, these functions are related by charge conjugation, and thus the
spinors $U$ and $V$ correspond to particle and anti-particle states.
In the presence of a matter background the charge conjugation symmetry
C is spontaneously broken.  The time-dependent phase factor in
eq. (\ref{Ansatz}) is introduced to restore this symmetry for the mode
functions $\phi^{(\lambda)}_{k,\pm}$; under charge conjugation the sign
of the momentum and of the reduced density $\rho$ are changed
similtanuously.

We choose initial conditions at $t \leq 0$ corresponding to a constant
matter background. The mode functions are then simply plane wave
solution with energies that can be read off from the mode equation
(\ref{mode}),
\begin{equation}
	\phi_{k,\pm}^{(\lambda)}(0) = 1, \quad
  	\dot{\phi}_{k,\pm}^{(\lambda)}(0) = \lambda E_{k,\pm}(0) 
	\quad {\rm with } \;
	E_{k,\pm}(0) = \sqrt{m^2 +k_{\rm eff, \pm}^2(0) }.
\label{ini}
\end{equation}
In the limit $\rho \to 0$, the spinors $U$ and $V$ become the usual
free field positive and negative frequency solutions. We normalize the
spinors $U_{{k},r}^{\dag}(0) U_{{k},s}(0) = V_{-{k},r}^{\dag}(0)
V_{-{k},s}(0) = \delta_{rs}$. This fixes the normalization constant in
eq. (\ref{Ansatz}), $N_{k,\pm}^{-2} = (E_{k,\pm}(0) - k_{\rm eff,\pm}
(0))^2 + m^2$.  The normalization properties are preserved by the
unitary time-evolution of the system.

The system --- a neutrino in the presence of weakly charged matter ---
is solved by solving the mode equations, with given initial conditions
and normalization.  The next question then is: ``What do the solutions
mean in terms of particle production?''

\section{Particle number}

The number operator constructed from the creation and annihilation
operators in the mode expansion, eq.~(\ref{exp}) is time-independent.
Clearly, this will not do to discuss particle production. Particle
production is associated with the time-dependence of the Hamiltonian.
One can introduce time-dependent creation and annihilation operators
that diagonalize the Hamiltonian instantaneously.  This defines a
physically motivated, time-dependent number operator.  Equivalently
one can describe particle production in terms of the mode
functions. The initial mode functions are positive and negative
frequency solutions to the mode equation.  However, due to the
time-dependence of the Hamiltonian, at later times they will in
general be a superposition of positive and negative frequencies, the
overlap being a measure of particle production. One finds for the
particle number
\beq
	{\mathcal N}^{P}_{k,s}(t) =|D^{P}_{k,s}|^2, 
\label{bog}
\eeq
where the so-called Bogoliubov coefficient $D^{P}_{k,s}= {U_{{\bf
k},\pm}}^{\dag}(0) V_{-{\bf k},\pm}(t)$ is (up to a phase)
\beq
	D^{P}_{k,\pm} 
	= {m}{N_{{\bf k},\pm}^{2}} \left[ i \dot{\phi}^{(-1)}_{k,s}(t) -
	\phi^{(-1)}_{k,s}(t) \left(E^{(-1)}_{k,\pm}(0) +
	\frac{1}{2}(\rho_\pm(0)- \rho_\pm(t)) \right)  \right ].
\label{Dpart}
\eeq
Both $U$ and $V$ are unitary matrices, and
thus the Boguliubov coefficient cannot exceed unity, in accordance
with the Pauli exclusion principle. Note further that particle
production goes to zero in the limit of a massless neutrino. This is
as expected, since in this limit the left and right handed sector
decouple, and if the Hamiltonian is diagonal at the initial time, it
remains so at all times.

\section{Results}

\begin{figure}[tb]
\begin{centering}
\includegraphics[height=5.3cm]{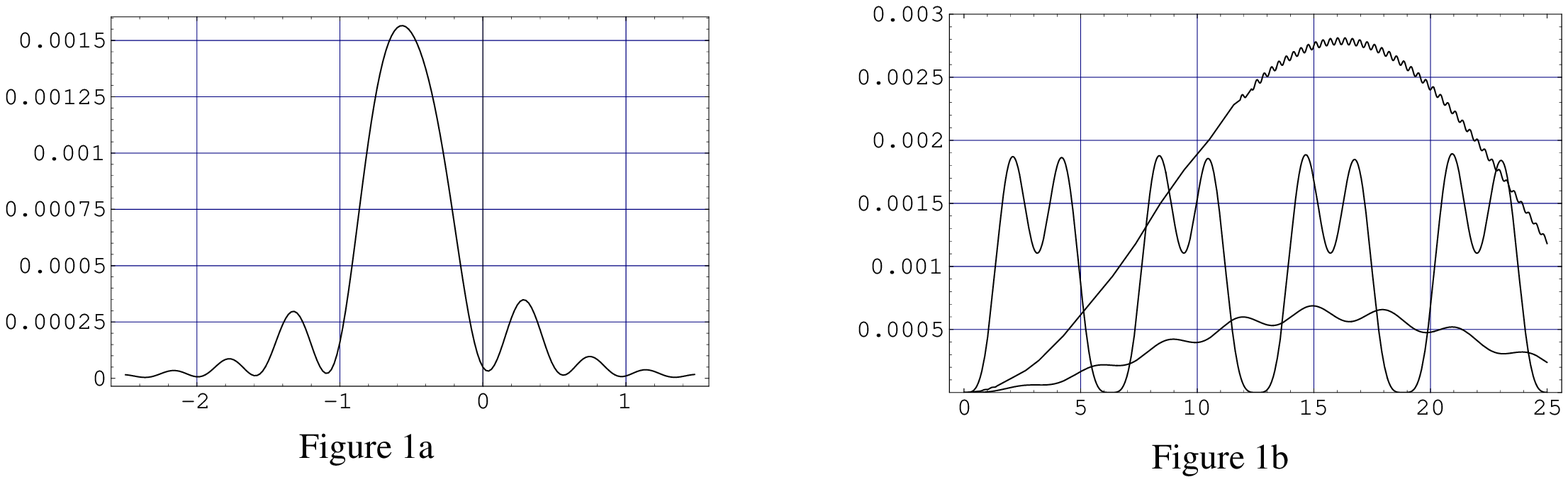}
\caption{Paricle production in an oscillating matter background $j^0 =
1 + 0.1 \sin(\omega t)$ and $\vec{j}=0$. Fig. 1a shows ${\mathcal N}_{{
k},+}$ vs.  momentum $k$ for $\omega=1$ and time $t=10$.  Fig. 1b
shows ${\mathcal N}_{{ k},+}$ as a function of time $t$.  The upper
curve corresponds to $\!(\omega,k) = (10,4.3)$ (close to resonance),
the middle curve to $(\omega,k) = (1,-0.5)$, and the lower curve to
$(\omega,k) = (0.1,-0.5)$. All quantities are in units of $m$.}
\end{centering}
\end{figure}

As an explicit example we have calculated numerically the neutrino
production in a matter background that is changing periodically with
time: $j^0 = \rho(t) = 1 + 0.1 \sin(\omega t)$ and $\vec{j}=0$
(all quantities in units of $m$).  The results are shown in figure 1a
and b.  Figure 1a shows that neutrino production is peaked where the
effective momentum $k_{\rm eff} = k + \rho/2$ is zero, and the
instantaneous energy appearing in the mode equation~(\ref{mode}) is
minimized.  The width of the excited band is set by the neutrino mass.
Figure 1b shows particle number as a function of time. It oscillates,
corresponding to creation and annihilation of neutrino pairs.  What is
left out of the picture is the diffusion of neutrinos as a result of
the finite size of the matter region, {\it i.e.}, neutrinos can escape
the production region before they are absorbed. If diffusion is
efficient, the production rate is of the order of the slope in fig 1b.

How effective is the neutrino production mechanism?  To get an idea,
we compare this process with another process in which fermions can be
produced, namely preheating.  In simple models of inflation the
inflaton field $I$ starts oscillates in a potential well at the end of
inflation. If the inflaton couples to a fermionic field, this leads to
copious production of fermions.  Particle production occurs in the
non-adiabatic regime, when the inflaton field goes through the minimum
of the potential well and $\dot{I} \gg I^2$.  As a result, the width
of the resonance band is set by the frequency of oscillations, which
can be much larger than the fermion mass. Furthermore, the
dimensionless effective inflaton-fermion coupling $\lambda \sim
I_0/\omega$ is large because of the large amplitude of oscillation
$I_0$. Fermion production is unsurpressed, and ${\mathcal N}^{P}_{k,s}
= {\mathcal O}(1)$.  Neutrino production on the other hand occurs in
the adiabatic regime, since for physically realistic values of
$\omega$ we have $\dot{\rho}/2 \ll E_{\rm ins}^2$ at all times (here
$E_{\rm ins} = \sqrt{m^2 + k_{\rm eff}^2}$ is the instantaneous energy
of the mode functions).  The width of the resonance is then set by the
neutrino mass, or by the size of the matter perturbation $\delta \rho$
if $m > \delta \rho$; both are at most the order of eV.  Moreover the
effective dimensionless coupling $\lambda = \delta \rho / m
\stackrel{<}{\sim} 1$, and particle number is surpressed by a factor
${\mathcal N}^{P}_{k,s} \propto \lambda^2$.

\section{Conclusions}

To summarize, neutrinos can be pair produced by a time-dependent
matter background.  Although neat, the phenomenon seems of little
practical importance.  Due to the smallness of all mass scales
involved, the amount of neutrinos produced is small.


\Acknowledgements

This work was supported in part by the US Department of Energy grant
DE-FG03-91ER40662, Task C, as well as by a Faculty Grant from UCLA Council
on Research.

\end{document}